\documentclass[12pt]{article}
\usepackage{amsmath}
\usepackage{authblk}
\usepackage{graphicx}

\title{Identification of Spikes in Time Series}
\author[1]{Dana E. Goin}
\author[1]{Jennifer Ahern}
\affil[1]{\footnotesize Division of Epidemiology, School of Public Health, University of California, Berkeley, California}
\date{\today}

\begin{document}

\maketitle 


\newpage
\section*{Abstract}
Identification of unexpectedly high values in a time series is useful for epidemiologists, economists, and other social scientists interested in the effect of an exposure spike on an outcome variable. However, the best method to identify spikes in time series is not known. This paper aims to fill this gap by testing the performance of several spike detection methods in a simulation setting. We created simulations parameterized by monthly violence rates in nine California cities that represented different series features, and randomly inserted spikes into the series. We then compared the ability to detect spikes of the following methods: ARIMA modeling, Kalman filtering and smoothing, wavelet modeling with soft thresholding, and an iterative outlier detection method. We varied the magnitude of spikes from 10-50\% of the mean rate over the study period and varied the number of spikes inserted from 1 to 10. We assessed performance of each method using sensitivity and specificity. The Kalman filtering and smoothing procedure had the best overall performance. We applied Kalman filtering and smoothing to the monthly violence rates in nine California cities and identified spikes in the rate over the 2005-2012 period.

\newpage
\listoftables
\listoffigures

\newpage
\section*{Introduction}
Identification of unexpectedly high values in a time series is useful for epidemiologists, economists, and other social scientists interested in the effect of an exposure spike on an outcome variable. Exposure spikes may be of interest when they are considered to be caused by something exogenous to the general patterning of the series, as in the case of income shocks and infant mortality \cite{baird2011aggregate}, thus strengthening the inference that can be drawn from the estimated effect \cite{humphreys2016changing}. Furthermore, spikes may be of interest when extreme increases in the exposure series are hypothesized to have disproportionate effects on the response compared to more usual disturbances from expected values or compared to corresponding decreases in the exposure. Previous studies have examined the effects of weather or economic spikes on outcomes as diverse as civil conflict, birth weight, and nutrition \cite{bhattacharya_heat_2003, jacob_dynamics_2007, miguel_economic_2004, margerison-zilko_maternal_2011, wilde2017effect}.

Community violence is another exposure that exhibits spikes. Exposure to community violence has been linked to stress-related health outcomes, including depression, asthma, cardiovascular disease, and birth weight \cite{kane2011ecology, brown2005traumatic, apter_exposure_2010, clark_witnessing_2008, koppensteiner2016violence, miller_witnessed_1999, martinez_nimh_1993, masi_neighborhood_2007, curry_pathways_2008}. However, many studies of community violence and health have suffered from structural confounding, in which the strong correlation of community violence with factors such as segregation, poverty, and unemployment means the effects are challenging to disentangle \cite{oakes_commentary:_2006, ahern_navigating_2013}. Examination of spikes in community violence offers advantages when between-community comparisons would suffer from structural confounding, because researchers can compare individuals within a community over time.  

To study the impacts of spikes in an exposure on health outcomes, spikes must be well characterized. In this study, we consider a spike to be an acute increase in the series followed by an immediate return to the underlying level of the series. This type of spike is described in the time series literature as an additive outlier \cite{cryer_time_2008}.  The best method for identification of spikes in time series is not known.  Many previous studies have defined spikes using pre-specified critical values above or below yearly averages \cite{bhattacharya_heat_2003, jacob_dynamics_2007, miguel_economic_2004, wilde2017effect}. However, such methods do not effectively account for underlying trends or autocorrelation. Some studies have used time series methods to identify spikes \cite{margerison-zilko_maternal_2011}, but a comparison of different methods has not been done. 

The aim of this paper is to describe several methods for identification of spikes in a time series, evaluate each method's performance in simulations, and illustrate an application to monthly violence rates in nine California cities. The code used for the simulations is provided in the supplemental materials for researchers interested in applying these methods to different exposure series. 

\section*{Simulation Study}
Assessment of outliers and extreme values in time series differs from the approaches used in non-time ordered data due to the potential for autocorrelation, trends, and cyclical patterns. We compared the following methods for identifying spikes: ARIMA modeling \cite{cryer_time_2008,metcalfe_introductory_2009}, Kalman filtering and smoothing \cite{de_jong_diagnosing_2000,durbin2012time}, wavelet modeling with soft thresholding \cite{nason_wavelet_2008}, and an iterative outlier detection method \cite{cryer_time_2008, chan2012tsa}. We selected these methods because they represent the most common time series modeling and outlier detection approaches. All methods were implemented using R packages available from CRAN. Additional details about each spike detection method are provided in the following sections. 

\subsection*{Summary of simulation}

To compare each method’s ability to identify spikes, we devised a collection of simulation studies parameterized based on violence rates in several California cities. The cities we selected had monthly violence rates that differed in terms of mean, variance, and autocorrelation. 

Each city’s simulation study included 1) simulating a series parameterized to be similar to city-level violence data, 2) adding a pre-specified magnitude and number of spikes to the series, 3) applying each detection method and calculating sensitivity and specificity of each method, and 4) iterating the procedure 1,000 times. 

We conducted simulations with a range of spike numbers and magnitudes. The magnitudes ranged from 10 to 50 percent increases over the average rate during the study time period. The number of spikes inserted into the series ranged from 1 to 10. These variations allowed us to assess the ability of each method to detect different magnitudes of spikes and to determine whether the number of spikes present in the series influenced the performance of each method.

\subsection*{Data description}
We selected the following cities in California for our study: Berkeley, Fresno, Oakland, Los Angeles, Richmond, Sacramento, San Diego, San Francisco, and Stockton. We selected these cities because they range in population size and their violence rates have a range of characteristics. 

Interpersonal violence totals by city were created by summing the total number of deaths and injuries attributable to assault or homicide from the emergency department records and patient discharge and inpatient hospitalization records from the Office of Statewide Health Planning and Development (OSHPD) and the death records from Vital Statistics. To estimate monthly rates, we divided the number of cases in each city by the estimated number of people living in each city in each month and multiplied by 100,000. The population denominators came from the intercensal and postcensal population estimates from the U.S. Census Bureau. These data capture all assaults severe enough to require an emergency department visit or hospital stay and all homicides of California residents during 2005-2012.

\subsection*{Simulation details}
Each city had its own simulation study. First, we fit the city's actual violence series with an ARIMA model whose parameters were selected by the Aikake Information Criteria (AIC). We then simulated from this model in order to capture general properties of the series (such as mean, variance, autocorrelation, and trend). 

\subsection*{Fitting initial ARIMA models}

 An ARMA model predicts current values of the response based on past values (autoregressive (AR) parameters) and innovations or past error values (moving average (MA) parameters). If differencing is required, the ARMA model is integrated and described as an ARIMA model. The standard equation of an ARIMA model is  
\begin{equation}
  y_t=\mu + \phi_1 y_{(t-1)}+\dots +\phi_p y_{(t-p)}-\theta_1 e_{(t-1)}- \dots - \theta_q e_{(t-q)}
\end{equation}

which is commonly expressed as 
\begin{equation}
  \phi(B)(1-B^d) y_t= \mu - \theta(B) \epsilon_t
\end{equation}

where B is the backshift operator (where $B^d y_t=y_{(t-d)}$), $\phi$ and $\theta$ are polynomials of order p and q, respectively, d is the amount of differencing, and $\epsilon$ is a noise process with assumed distribution $N(\theta,\sigma^2)$.

The order of the AR portion is usually referred to as p and the order of the MA portion is usually referred to as q. An ARIMA(p,d,q) model has AR degree p, is differenced d times, and has MA degree q. We used \textit{auto.arima} function from the \textit{forecast} package in R, which uses AIC to select the model order \cite{hyndman2007automatic}. The \textit{auto.arima} function uses a unit root test with a null hypothesis of no unit root to pick the amount of differencing required. Once the proper level of differencing is determined, a stepwise algorithm is used to select the model order, which is described in detail in the documentation for the \textit{forecast} package \cite{hyndman2007automatic}. The model coefficients are estimated using maximum likelihood. The model parameters selected for each city are listed in Table \ref{table:arimaparam}.

\subsection*{Inserting spikes}
We simulated from each city's fitted ARIMA model and inserted spikes randomly into the series, with all time points equally likely to receive a spike. The number and magnitude of spikes were varied from 1 to 10 and from 10\% to 50\% of the mean violence rate, respectively. Each combination of spike magnitudes and numbers were run as separate simulation studies and replicated 1,000 times.

\subsection*{Applying spike detection methods}
A spike was considered correctly identified when a method identified a spike in violence at the time point where one was inserted by our algorithm. Any spike identified at a time point where we had not inserted a spike was considered to be incorrectly identified. We summarized the performance of each method using sensitivity and specificity. Descriptions of each method follow.

\subsection*{ARIMA}
As described previously, an ARIMA model predicts the current value of the series using a linear combination of past values and innovations. The innovations are assumed to be $N(0,\hat{\sigma^2})$. When using the ARIMA model to identify violence spikes, we fit the simulated series with an ARIMA model selected via AIC using the \textit{auto.arima} package \cite{hyndman2007automatic}. We then calculated the residuals and their standard deviation. Any time point with a residual value greater than two times the standard deviation of the residuals was identified as a spike. 

\subsection*{Kalman filter and smoother}

The Kalman filter is a recursive data processing algorithm most famously used in engineering problems to predict trajectories based on position and velocity \cite{faragher2012understanding, durbin2012time}. It uses the observed values of a series to update its predictions, resulting in a weighted average of the predicted and observed values where the weights are determined by the uncertainty around each. The Kalman filter uses a state space approach to time series modeling, and attempts to model a latent state that is unobserved but for which there are recorded measurements related to the state at discrete points in time. In this way, the state space approach is similar to a hidden Markov model in which the state space of the latent variable is continuous and the latent and observed variables are assumed to be Gaussian distributed. 

The state space approach requires two equations, called the state equation and the observation equation. For details of the equations and derivations in this section, see \cite{durbin2012time}. 

The equation for the state of the system is
\begin{equation}
  \alpha_t = T_t\alpha_{t-1}  + R_t\eta_t
\end{equation}

where $\alpha_t$ is the state of the system at time $t$, $T_t$ is the transition matrix, which applies characteristics of the system at time $t-1$ to generate a prediction of the state at time $t$, and $\eta_t$ is a vector of error terms with assumed distribution $N(0,Q_t)$. In our example, the matrix $R_t$ is the identity matrix. 

There is also an observation equation for the system, defined as 
\begin{equation}
  y_t = Z_t\alpha_t + \epsilon_t
\end{equation}

where $y_t$ is the measured value at time $t$, $\alpha_t$ is the true state at time $t$, $Z_t$ is the matrix that maps the state to the measured value, and $\epsilon_t$ is a vector of measurement error terms assumed to be distributed $N(0, H_t)$. 

The following matrices must be specified or estimated: 
\begin{itemize}
	\item $T_t$,  the state-transition matrix, which maps the state at time $t-1$ to the state at time $t$
\item  $Z_t$, the observation matrix which maps the true state to the observed state at time $t$
\item $Q_t$, the covariance of the state process at time $t$
\item $H_t$, the covariance of the observation measurement at time $t$
\end{itemize}

The Kalman filter itself consists of the following recursion equations: 
\begin{equation}
	v_t = y_t – Z_ta_t
\end{equation}
where $v_t$ is the residual of the observed minus predicted value of the (latent) state, 
\begin{equation}
a_t  = E(\alpha_t | y1, \dots, y_t-1)
\end{equation}
 where $a_t$ is the expected value of the state given past observations, 
\begin{equation}
	a_{t|t} = a_t + P_tZ_t’F_t^{-1}v_t
\end{equation}
where $a_{t|t}$ is the expected value of the state given the past and present observations, 
\begin{equation}
P_t = var(\alpha_t | y_1, \dots, y_{t-1})
\end{equation}
where $P_t$ is the variance of the state given past observations, 
\begin{equation}
	F_t = var(v_t | y_1, \dots, y_{t-1}) = Z_tP_tZ_t’ + H_t
\end{equation}
where $F_t$ is the variance of the innovations, given past observations  
\begin{equation}
	a_{t+1}  = T_ta_{t|t} = T_ta_t + K_tv_t
\end{equation}
where $a_{t+1}$ is the prediction for the value of the state at the next time point given past observations, 
\begin{equation}
	K_t = T_tP_tZ_t’F_t^{-1}
\end{equation}
where $K_t$, referred to as the Kalman gain, represents the change in the estimate of the state at time $t$ after incorporating the information from the measurement, 
\begin{equation}
P_{t|t} = P_t – P_tZ_t’F_t^{-1}Z_tP_t
\end{equation}
where $P_{t|t}$ is  the variance of the state given past and present observations, and 
\begin{equation}
P_{t+1} =  T_tP_t(T_t – K_tZ_t)’ + R_tQ_tR_t’
\end{equation}
where $P_{t+1}$ is the variance of the prediction for the state at the next time point, given past observations. 

The smoothing algorithm combines the results from the Kalman filtering and does backward recursion to create new smoothed estimates for the state. This is done for each time $t$ by combining both the estimate of the state at time $t$ from the filtering process and the residuals at times $>t$. The state smoothing recursion equations are:

\begin{equation}
\begin{split}
	\hat{\alpha_t} &= a_t + P_tr_{t-1}  \\
r_{t-1} &= Z_t’ F_t^{-1}v_t + L_tr_t \\
N_{t-1} &= Z_t’F_t^{-1}Z_t + L_t’N_tL_t  \\
V_t &= P_t- P_tN_{t-1}P_t \\
\end{split}
\end{equation}

where $r_t$ is a weighted combination of the innovations that occur after time $t-1$ and $L_t =T_t – K_t = T_t-T_tP_tZ_t’F_t^{-1}$. 

We used the \textit{KFAS} package in R to apply Kalman filtering and smoothing to the data \cite{helske2014kfas}, which uses the algorithms from \cite{durbin2012time}. For clarity, we have tried to stay as consistent as possible with the notation from these sources. 

For our example, the latent state $\alpha_t$ is the violence level at time $t$ and the observed data is the measured violence rate $y_t$ at time $t$. To apply Kalman filtering and smoothing, we first fit an ARIMA model to the series, which was selected using the AIC, and transformed it into a state-space model formulation. The recursive filtering and smoothing algorithms described above were applied to the data, creating smoothed predictions for the violence level at each time point. Finally, we calculated residuals and identified spikes as any value greater than two times the standard deviation of the residuals.

\subsection*{Wavelets}
Wavelets are functions that oscillate around zero and satisfy $\int_{-\infty}^{\infty} \psi(x)dx=0$ \cite{nason_wavelet_2008}. The wavelet transform is similar to a Fourier transform, in that for a given function $f(x)$, we can represent the function as a sum of orthonormal functions \cite{nason199discrete}. However, instead of expressing the function as a sum of sine and cosine or complex exponential terms like in a Fourier series, wavelet analysis uses functions that are dilations and translations of a function called a mother wavelet $\psi$ \cite{nason_wavelet_2008}. 

For example, we can define a set of wavelets 
  \begin{equation}
    \psi_{j,k}(x) = 2^{j/2}\psi(2^jx-k)
  \end{equation}
where $j$ and $k$ are integers, which have the effect of dilating $\psi$ by a factor of $2^j$ and translating by $2^{-j}k$, and $2^{j/2}$ is a normalizing constant. These wavelets form an orthonormal set, which implies that a function $f(x)$ can be decomposed into a generalized Fourier series
  \begin{equation}
    f(x) = \sum_{j=-\infty}^{\infty} \sum_{k=-\infty}^{\infty} b_{j,k}\psi_{j,k}(x)
  \end{equation}
where $j$ and $k$ are integers that index the dilation number ($j$) and the translation number ($k$) \cite{nason_wavelet_2008}. The wavelet coefficients $b_{j,k}$ are the inner product of $f(x)$ and $\psi_{j,k}(x)$ such that
\begin{equation}
  b_{j,k} = \int_{-\infty}^{\infty}f(x)\psi_{j,k}(x)dx
\end{equation}
 
The wavelet transform is the process of representing the function in terms of the wavelets and their coefficients. The goal of using a wavelet transform is to calculate the coefficients that allow us to approximate the series. Thresholding the coefficients allows for different levels of smoothing. Soft thresholding shrinks the coefficients around the threshold. For a more complete description of wavelets, their properties and usage, see \cite{nason_wavelet_2008, nason199discrete}.

We used the R package \textit{wavethresh} to apply a wavelet transform to the data \cite{nason2010wavethresh, nason_wavelet_2008}. First a discrete wavelet transform was performed according to Mallat's pyramidal algorithm, and the coefficients were thresholded using a soft threshold. The wavelets were reconstructed by applying the inverse discrete wavelet transform. We calculated residuals and identified spikes as values greater than two times the standard deviation of the residuals. 

\subsection*{Outlier detection}
We also tested the performance of an iterative model fitting and outlier detection procedure based on \cite{chang1988estimation, tsay_outliers_1988}. 

The outlier detection process goes as follows: 
\begin{enumerate}
	\item Fit an ARIMA model to the observed time series.
\item Calculate the residuals and the variance of the residuals.
\item Compute the likelihood ratio test statistic $\lambda_t$ for an additive outlier at time $t$ 
	\begin{equation}
	\lambda_t = \frac{\hat{\omega_t}}{\hat{\rho_t}\hat{\sigma}}
	\end{equation}
		where 
		\begin{equation}
			\hat{\omega_t} = \hat{\rho_t^2}(1 - \sum_{i=1}^{n-t}\hat{\pi_i}\epsilon_t)
		\end{equation}
		and 
	\begin{equation}
		\hat{\rho_t^2} = (1 - \sum_{i=1}^{n-t}\hat{\pi_i^2})^{-1}
	\end{equation}
The $\pi$-weights are functions of the estimated coefficients of the ARIMA model, and can be expressed as $\pi(B) = \frac{\phi(B)}{\theta(B)}$. These weights can also be derived by representing the model as a recursive AR model.  
	\item Find the maximum in absolute value of the test statistics, and compare to a pre-specified critical value. 
	\item If the maximum exceeds the critical value, an outlier has been detected. Remove its effect by defining a new residual for the relevant time points and recalculate the residual variance. The adjusted residuals should have the form
	\begin{equation}
		\tilde{\epsilon_t} = \hat{\epsilon_t} - \hat{\omega}\hat{\pi}(B)I(t=T) \text{ for } t \geq T
	\end{equation}
where $I()$ is the indicator function. 
	\item Calculate new test statistics using the modified residuals and residual variance. 
	\item Continue until no more outliers are identified. 
	\item Assume the outliers identified above are known outlier times. Estimate the model parameters and the $\omega$ at each outlier time. 
	\item Use the parameter estimates from the models with assumed known outlier times and begin the outlier detection process again. Continue until no outliers are found. 
	\item If additional outliers are detected, re-estimate the parameters, incorporating the new outliers into the “known” outlier times. 
	\item Once no more outliers are identified, the procedure is complete. The locations of outliers (if any) in the time series have been identified and model parameters that exclude the effects of outliers have been estimated. 
\end{enumerate}

We used the \textit{detectAO} function in the \textit{TSA} R package to implement this outlier detection procedure \cite{chan2012tsa}.

\subsection*{Hypothesized performance across methods}
The methods we selected have different strengths and are thus likely to perform best under different scenarios. We expect ARIMA modeling to perform best when there are few spikes and the data have simple parameterizations. When there are many spikes, the coefficients fit by the ARIMA model may be biased due to the spikes, and spike detection may suffer. While the Kalman filter and smoother assumes linear equations and Gaussian errors like the ARIMA models, the Kalman filter is more adaptive than ARIMA because of the extra uncertainty it incorporates in the measurement equation and the updating step. Therefore, we expect the Kalman filter to out-perform the ARIMA model in situations of high variance. However, because of the updating step, we expect the Kalman filter to have worse specificity as it may incorrectly adjust toward lower spike levels, obscuring their effects. In contrast to the ARIMA and Kalman methods, wavelets are useful when data have localized patterns, non-linearities, and discontinuities. Therefore, we expect these methods to perform well when the series are not well characterized by ARIMA models or when there are many spikes in a series. Outlier detection methods may perform similarly to the ARIMA and Kalman methods in situations with large shocks, but may have better performance with small shocks. The iterative procedure may capture small spikes better than other methods because they remove large spikes before searching for smaller magnitude spikes.

\section*{Results}
The performance of all methods varied substantially by city and by series characteristics, although the patterns of performance were similar across magnitudes of spikes (Table \ref{table:sensitivity}, Table \ref{table:specificity}). Performance improved across all methods with increasing spike magnitude, as expected (Table \ref{table:sensitivity50}, \ref{table:specificity50}). In general, the places in which methods had higher overall performance also tended to have a higher ratio of mean to standard deviation.

Overall, the Kalman method had the highest sensitivity and the outlier detection method had the highest specificity. However, the outlier detection method was by far the worst performer in terms of sensitivity. The Kalman method had both high sensitivity and specificity, and was consistently the best performer with an average sensitivity of 89.98\% and an average specificity of 99.41\% across cities in simulations with 1 to 10 spikes inserted and spikes with magnitude 50\% increase over the series mean (Table \ref{table:sensitivity50}).  Sensitivity was lower for spikes of lower magnitude, which makes sense given these are much more likely to blend into the background variation. Averaging over all simulation scenarios, in which we included spikes of all magnitudes (10\%, 20\%, 30\%, 40\%, and 50\% increase over the series mean) and numbers of spikes from 1-10, the Kalman filter and smoother had an average sensitivity of 63.40\% and specificity of 98.49\% across cities (Table \ref{table:sensitivity}). The ARIMA method was a close second best performer, with an average sensitivity of 61.05\% and specificity of 98.28\% across all simulations. 

All methods performed worst in the simulation parameterized to be similar to Berkeley, which is likely due to the high relative variance, which made distinguishing spikes from background variation difficult. However, the Kalman filter method was still the best performer, with an average sensitivity of 40.07\% and specificity of 97.38\% for all magnitude spikes and across simulations with 1 to 10 spikes (Table \ref{table:sensitivity}, Table \ref{table:specificity}). Among the highest magnitude spikes (50\% increase over Berkeley series mean), the Kalman method had average sensitivity of 76.13\% and specificity 98.37\% (Table \ref{table:sensitivity50}, Table \ref{table:specificity50}).

\begin{figure}
    \centering
    \includegraphics[width=0.9\textwidth]{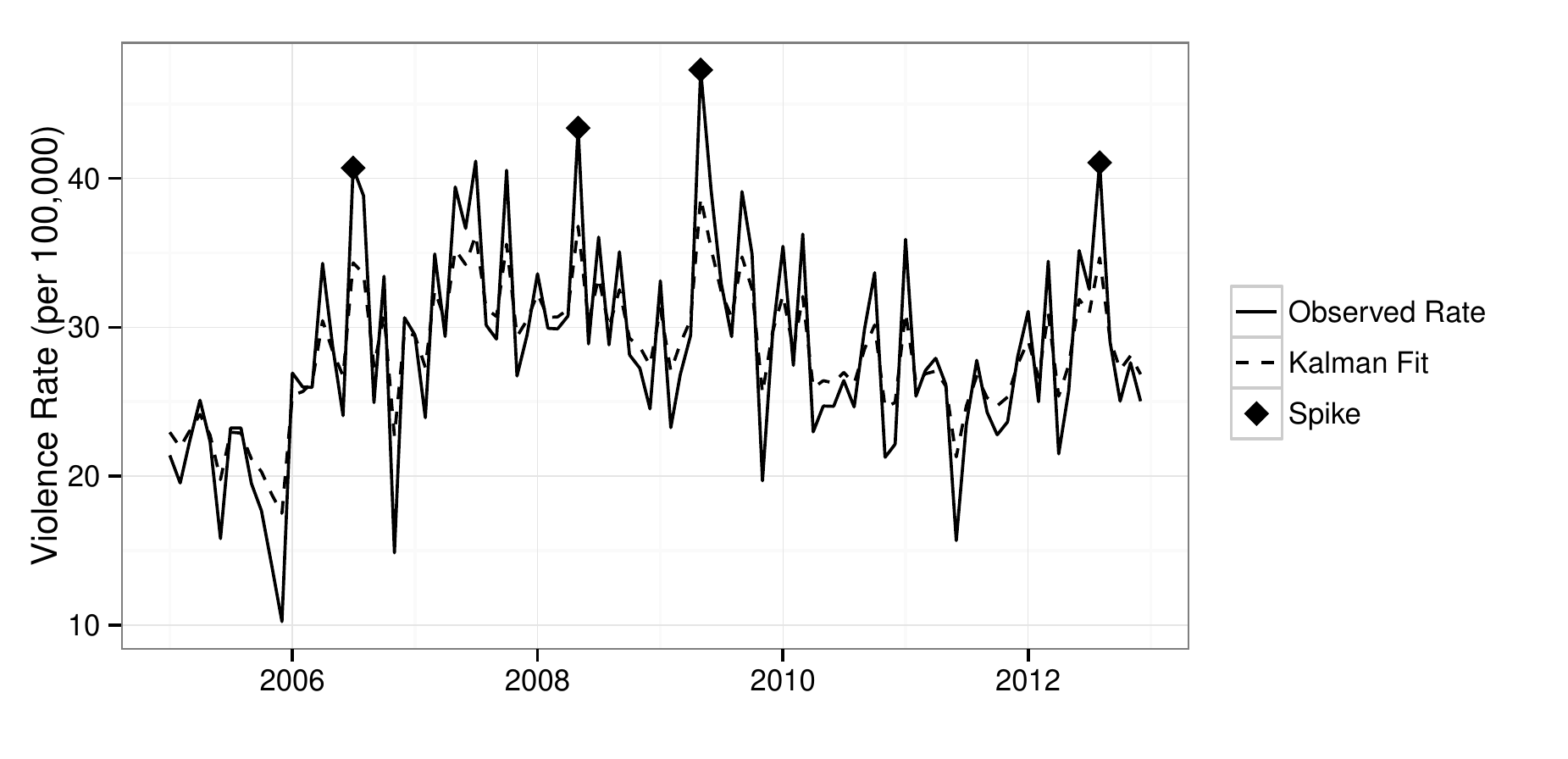}
    \caption{Kalman filter and smoother fit and spikes identified in month violence rate in Berkeley, CA from 2005-2012.}
    \label{fig:berkkfit}
\end{figure}

\begin{figure}
    \centering
    \includegraphics[width=0.9\textwidth]{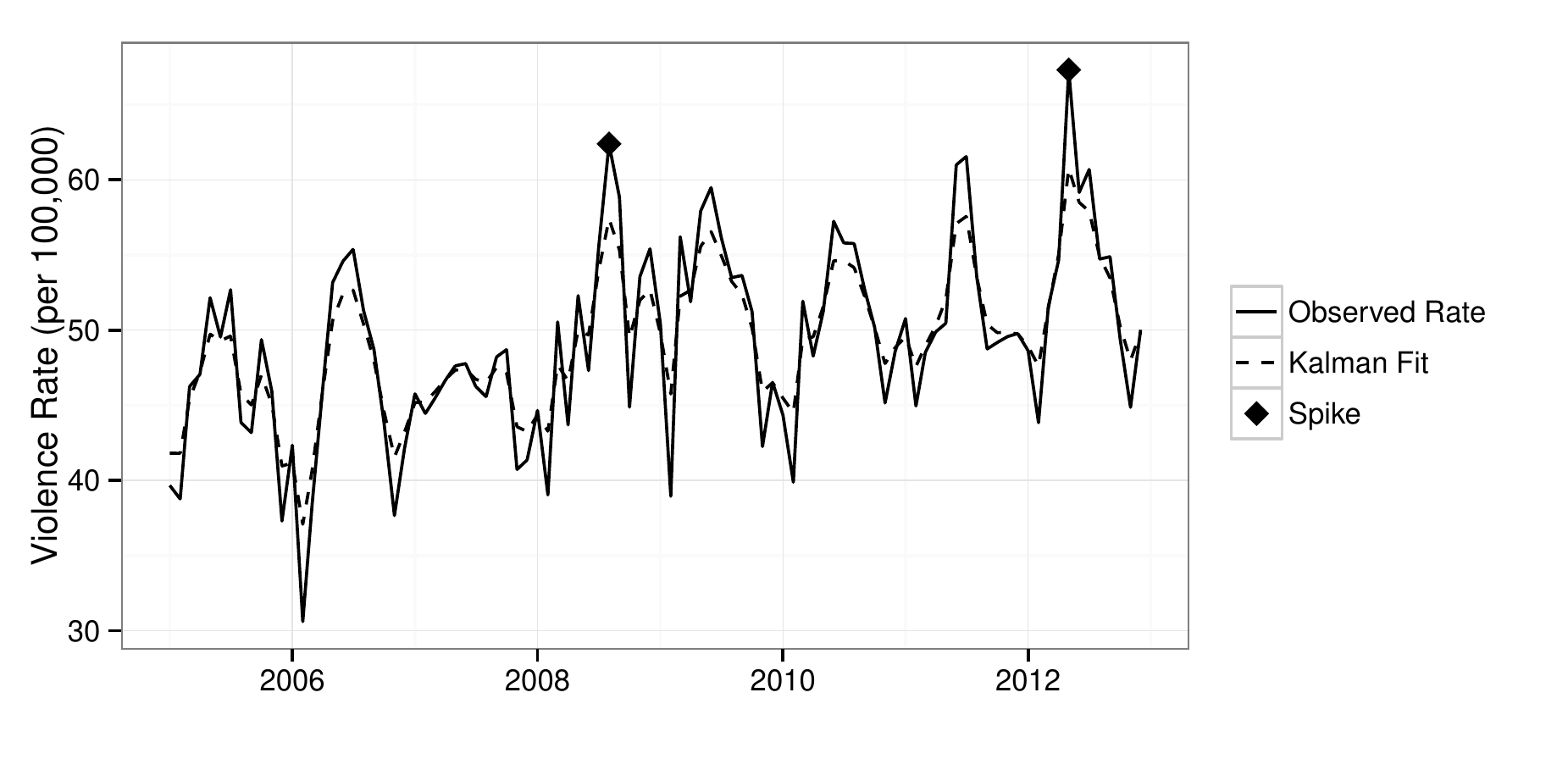}
    \caption{Kalman filter and smoother fit and spikes identified in month violence rate in Fresno, CA from 2005-2012.}
    \label{fig:freskfit}
\end{figure}

\begin{figure}
    \centering
    \includegraphics[width=0.9\textwidth]{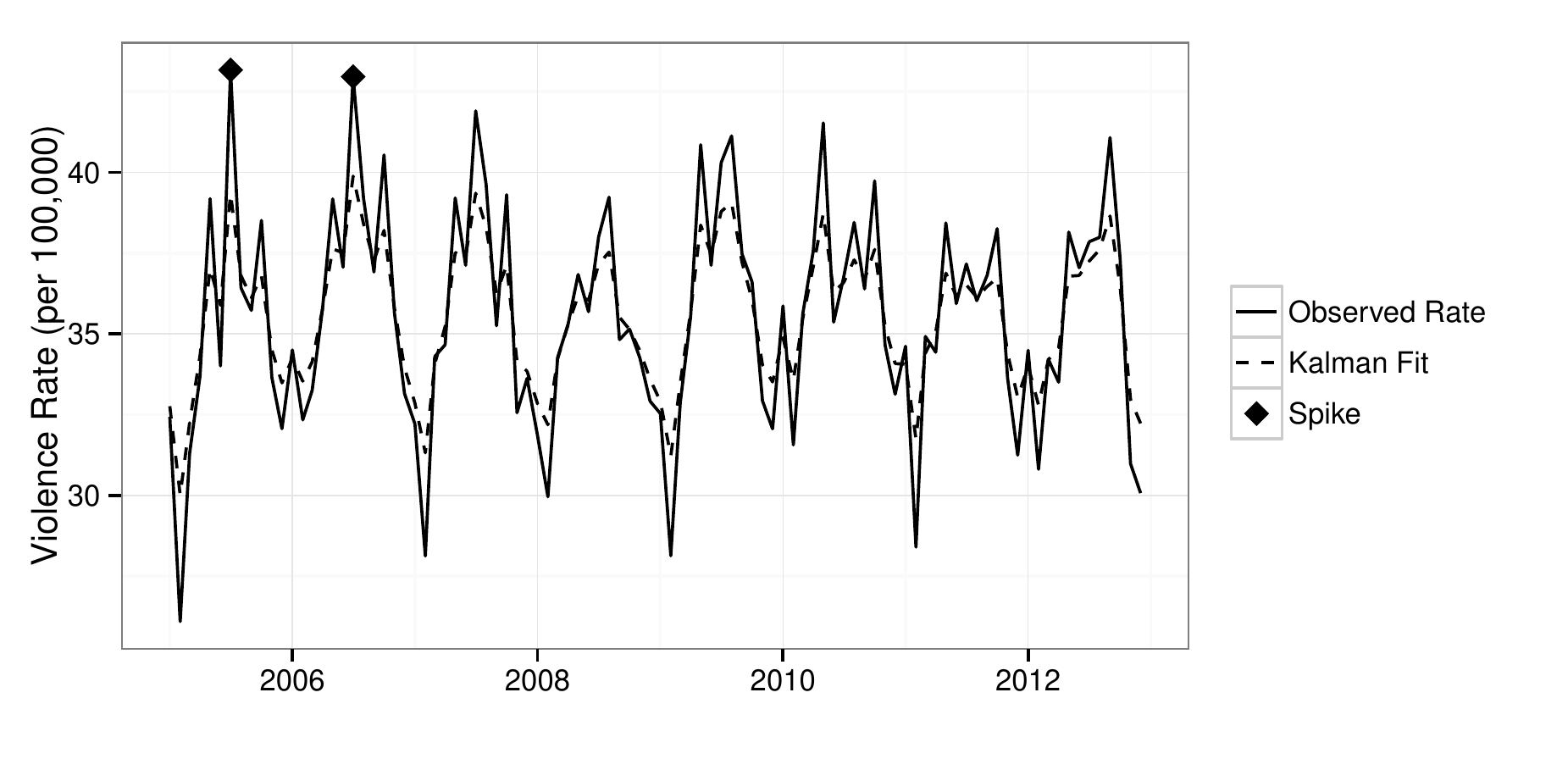}
    \caption{Kalman filter and smoother fit and spikes identified in month violence rate in Los Angeles, CA from 2005-2012.}
    \label{fig:lakfit}
\end{figure}

\begin{figure}
    \centering
    \includegraphics[width=0.9\textwidth]{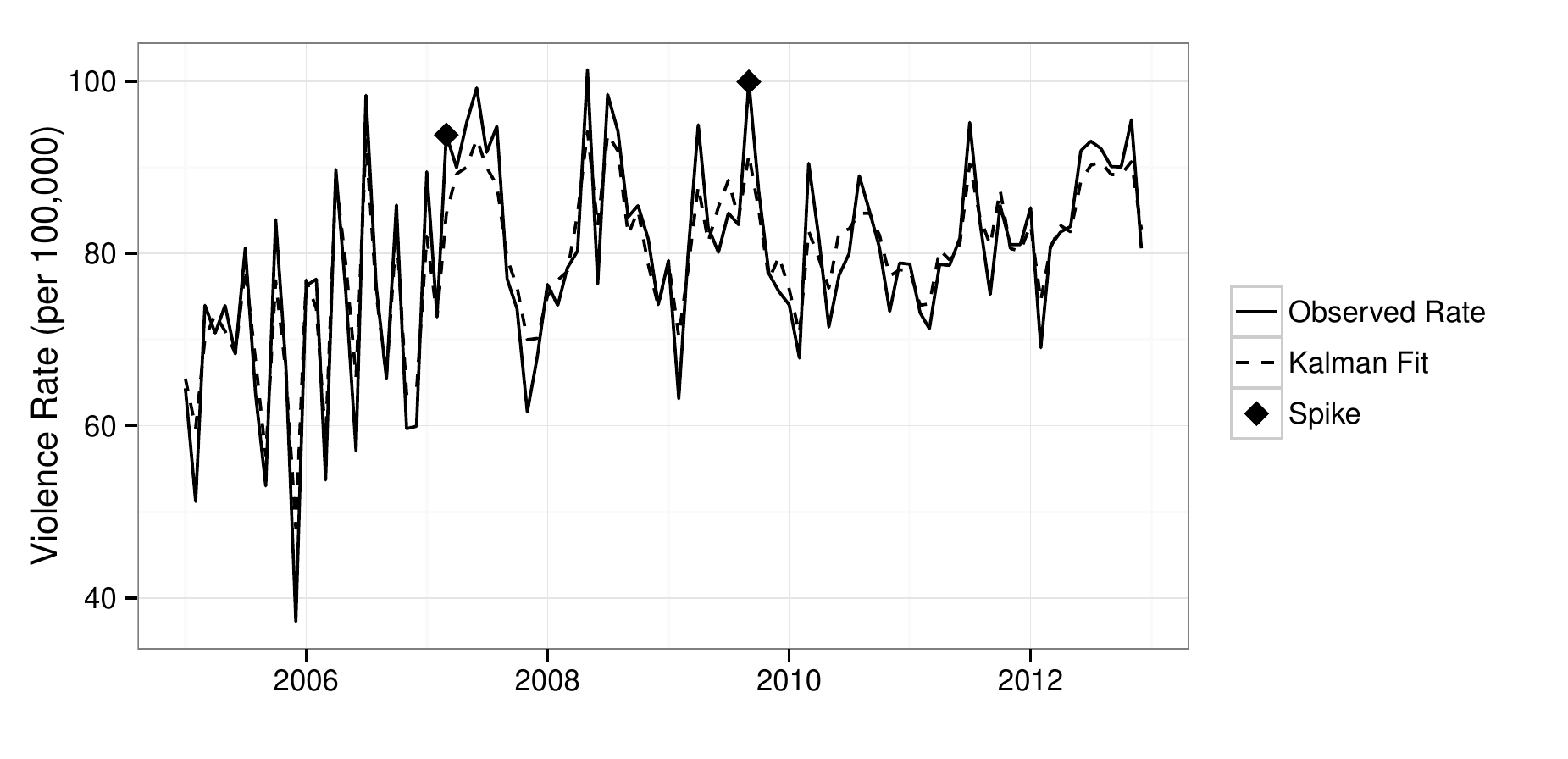}
    \caption{Kalman filter and smoother fit and spikes identified in month violence rate in Oakland, CA from 2005-2012.}
    \label{fig:oakkfit}
\end{figure}

\begin{figure}
    \centering
    \includegraphics[width=0.9\textwidth]{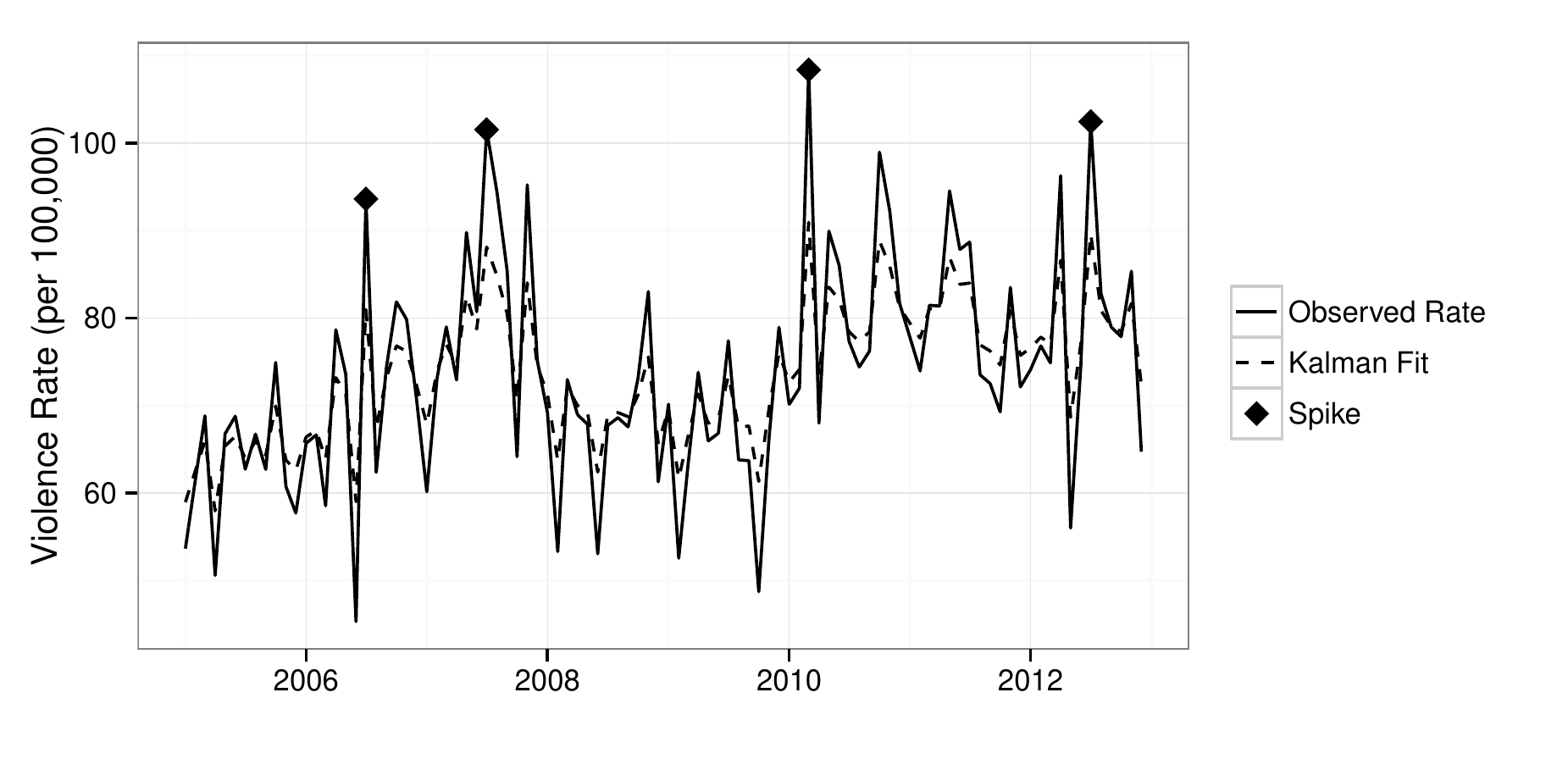}
    \caption{Kalman filter and smoother fit and spikes identified in month violence rate in Richmond, CA from 2005-2012.}
    \label{fig:richkfit}
\end{figure}

\begin{figure}
    \centering
    \includegraphics[width=0.9\textwidth]{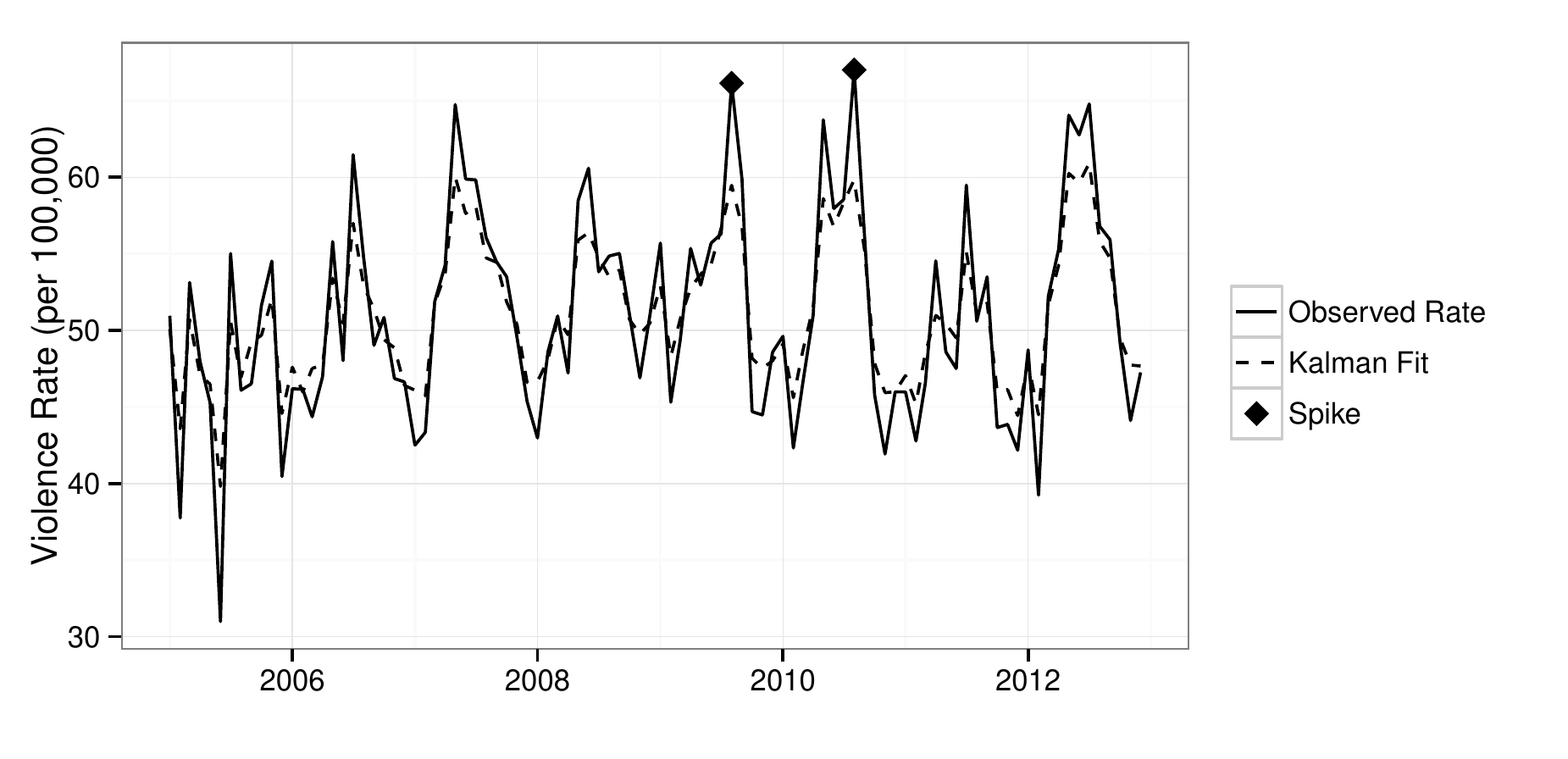}
    \caption{Kalman filter and smoother fit and spikes identified in month violence rate in Sacramento, CA from 2005-2012.}
    \label{fig:sackfit}
\end{figure}

\begin{figure}
    \centering
    \includegraphics[width=0.9\textwidth]{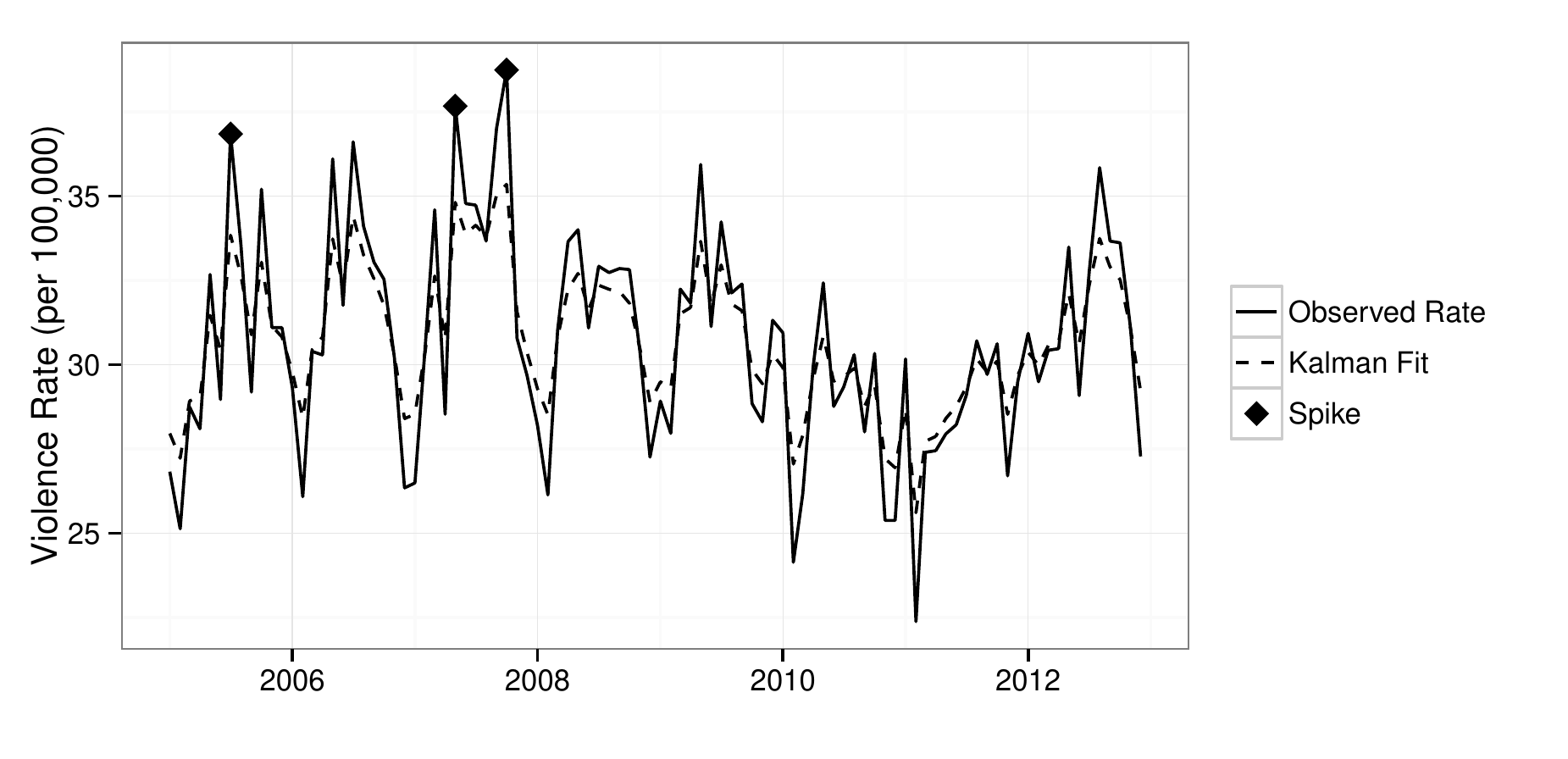}
    \caption{Kalman filter and smoother fit and spikes identified in month violence rate in San Diego, CA from 2005-2012.}
    \label{fig:sdkfit}
\end{figure}

\begin{figure}
    \centering
    \includegraphics[width=0.9\textwidth]{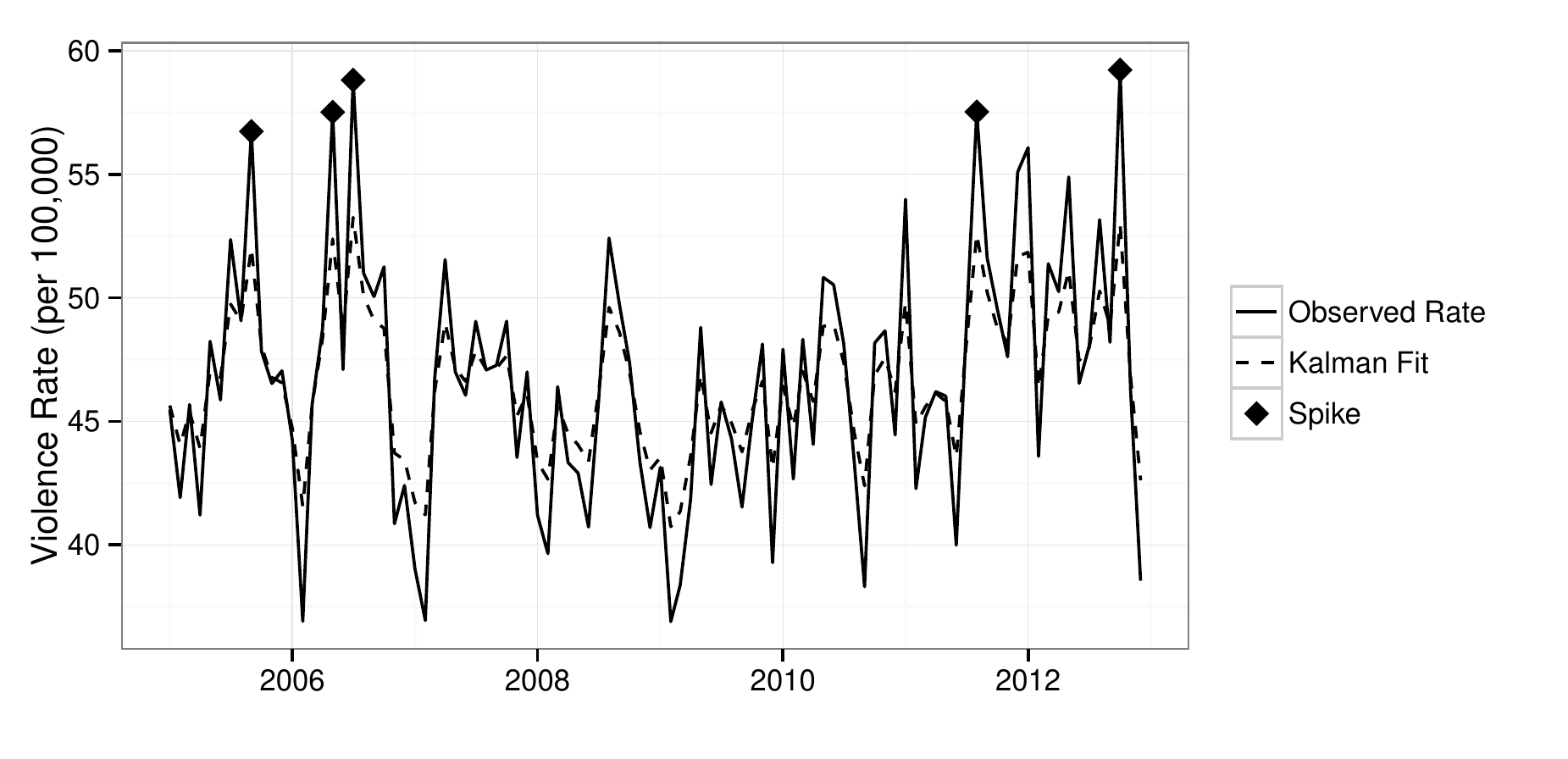}
    \caption{Kalman filter and smoother fit and spikes identified in month violence rate in San Francisco, CA from 2005-2012.}
    \label{fig:sfkfit}
\end{figure}

\begin{figure}
    \centering
    \includegraphics[width=0.9\textwidth]{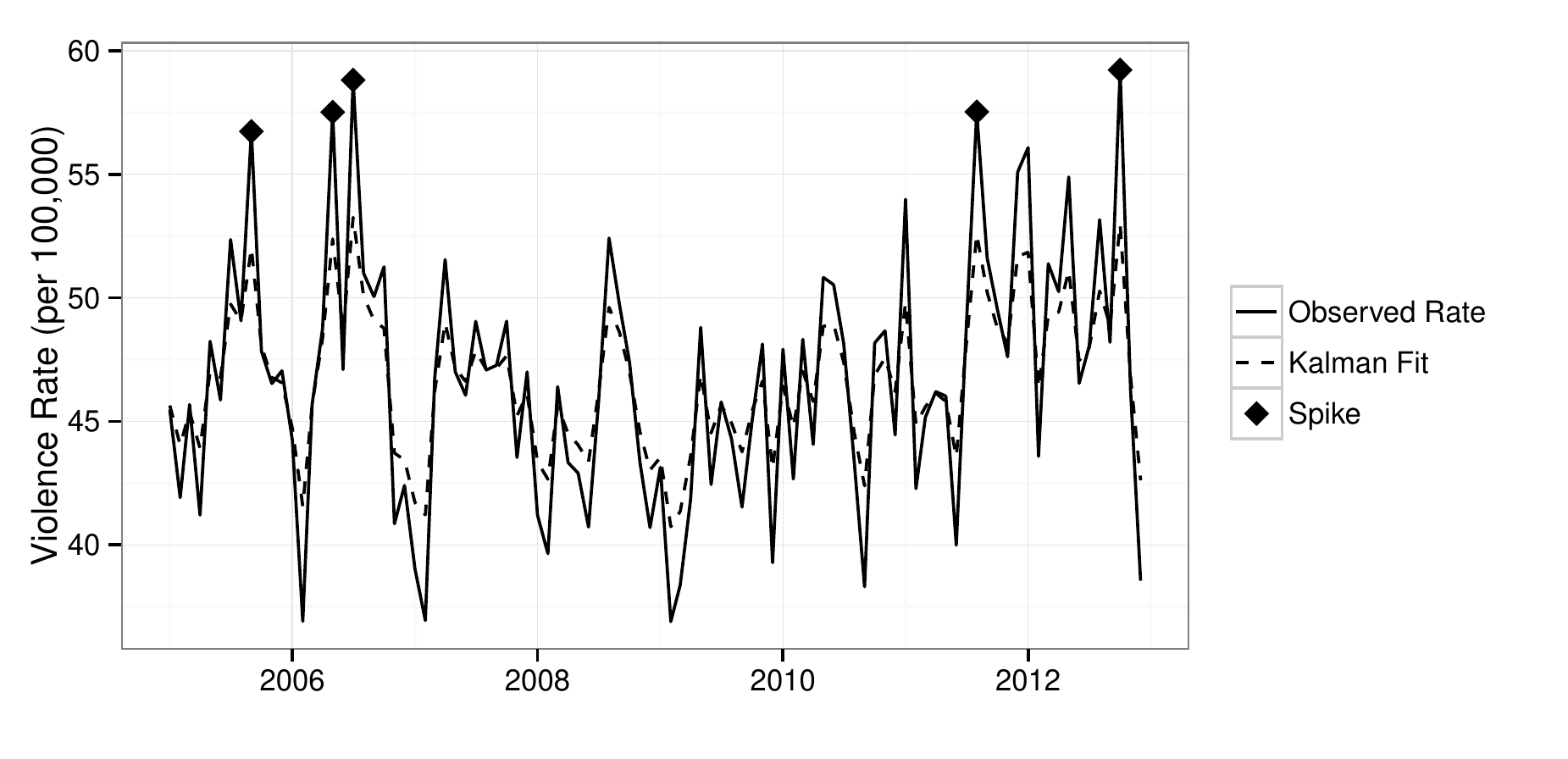}
    \caption{Kalman filter and smoother fit and spikes identified in month violence rate in Stockton, CA from 2005-2012.}
    \label{fig:stockkfit}
\end{figure}

We applied the Kalman method, the consistent best performer, to the true violence series for each of the nine cities to illustrate application. We identified several violence spikes in each city (Figures \ref{fig:berkkfit}, \ref{fig:freskfit}, \ref{fig:lakfit}, \ref{fig:oakkfit}, \ref{fig:richkfit}, \ref{fig:sackfit}, \ref{fig:sdkfit}, \ref{fig:sfkfit}, \ref{fig:stockkfit}). The months with spikes detected are listed in Table \ref{table:spikes}.

\section*{Discussion}

This work contrasted four methods that can be used to identify spikes in time series. We found that applying a Kalman filter and smoother and identifying values whose residuals were greater than two standard deviations from the predicted value was the best performing method in simulations parameterized based on violence time series from nine cities. The features of the simulations varied with respect to the mean, variance, autocorrelation, and trend, suggesting that across these features Kalman consistently does best. However, series of other data or violence series substantially different on these features might lead to different relative performance. The code provided in the supplemental material facilitates extension of this work to other series of interest that may have different features.

Our results are in accordance with our hypotheses that the Kalman filter and smoother would outperform the ARIMA when there are high magnitude spikes. However, while we thought the Kalman method would not perform as well with low magnitude spikes, it was the best performer for every magnitude and number of spikes. This method is easily implemented via the \textit{KFAS} package in R \cite{helske2014kfas}. Using this method, we found at least one spike in the monthly violence rate for each city. While the Kalman filter method was very successful in detecting large spikes, it inconsistently identified smaller magnitude spikes. 

There are some limitations to our approach. It is possible that simulating from ARIMA models to characterize the violence series does not completely capture the true underlying violence distributions. This may bias our results toward methods that rely on series well-described by ARIMA models. However, we would be more concerned about this if the ARIMA method were performing best in each scenario. While the Kalman filter also assumes linear functions and Gaussian errors, it incorporates a model for measurement error, which improves its ability to differentiate between normal variance and spikes in the series. 

It is also likely there are spikes in the violence series used to parameterize the simulations. This may affect the assumed data generating distribution for each city. However, we expect any bias in the parameter estimates will not affect the overall behavior of the simulated series. Repeating each simulation 1,000 times should incorporate sufficient variability around the true data generating mechanism so that small bias in the parameters should not affect the results. 

In order to assess how performance changed based on the critical values specified as part of each detection method, we also ran simulations increasing the threshold values. For the Kalman filter and smoother, ARIMA, and wavelet methods, this meant using a critical value of 2.5 standard deviations of the residuals rather than 2. There was a small decrease in sensitivity and increase in specificity, but the results were not substantively different.

This simulation study compared the performance of several methods for identifying spikes in time series. We generated separate simulation studies parameterized based on the violence rates in nine California cities that illustrated different series characteristics to assess performance across types of series. We varied the magnitude and the number of spikes inserted to the simulated series to explore how results varied in different scenarios that are plausible in real data. We found that applying a Kalman filter and smoother and identifying spikes as values above two standard deviations of the residuals consistently performed best in terms of sensitivity and specificity. The performance was best in cities that had low relative variance. Kalman filtering and smoothing is straightforward to apply using standard statistical software and should be considered by other researchers interested in the effect of exposure spikes on a response.

\newpage
\bibliographystyle{unsrt}
\bibliography{references}

\newpage
\begin{table}[ht]
\centering
\begin{tabular}{p{2.5cm}p{3.5cm}p{6cm}p{1cm}p{1cm}}
  \hline 
  \textbf{City} & \textbf{ARIMA model} & \textbf{Parameterization} & \textbf{Mean} & \textbf{SD}  \\ 
  \hline
   Oakland & ARIMA(4,1,2) &$(1-B)y_t = 0.8481y_{t-1} - 0.1436y_{t-2} + 0.3572y_{t-3} - 0.6178y_{t-4} + \epsilon_t -1.6616\epsilon_{t-1} + 0.7814\epsilon_{t-2}$  & 79.28 & 11.79   \\ 
   San Diego & ARIMA(2,0,0) & $y_t = 0.3605y_{t-1} + 0.1875y_{t-1} + \epsilon_t $& 30.72 & 3.20   \\ 
   Los Angeles & ARIMA(1,0,0) & $y_t = 0.436y_{t-1} + \epsilon_t $& 35.53 & 3.40  \\ 
   Sacramento & ARIMA(2,0,1) & $y_t = -0.465y_{t-1} + 0.5180\epsilon_{t-2} + \epsilon_t + 0.9315\epsilon_{t-1}$ & 50.99 & 6.85   \\ 
   Fresno & ARIMA(1,1,1) & $(1-B)y_t = 0.5306y_{t-1} + \epsilon_t - 0.9504\epsilon_{t-1}$ & 49.20 & 6.38   \\ 
   San Francisco & ARIMA(1,0,0) & $y_t = 0.3151y_{t-1} + \epsilon_t$ & 46.65 & 5.01  \\ 
   Stockton & ARIMA(1,0,0) & $y_t = 0.3440y_{t-1} + \epsilon_t$ & 54.58 & 7.90 \\ 
   Richmond & ARIMA(0,1,1) & $(1-B)y_t = \epsilon_t - 0.7940\epsilon_{t-1}$ & 73.59 & 12.49  \\ 
   Berkeley & ARIMA(1,0,1) & $y_t = 0.8762y_{t-1} + \epsilon_t -0.6531\epsilon_{t-1}$ & 28.41 & 6.84  \\ 
   \hline
\end{tabular}
\caption{ARIMA models and parameters by city}
\label{table:arimaparam}

\end{table}

\newpage
\begin{table}[ht]
\centering
\begin{tabular}{p{2.5cm}p{2cm}p{2cm}p{2cm}p{2cm}}
  \hline 
  \textbf{City} &           & \textbf{Sensitivity}  &           &  \\ 
                &  ARIMA    & Kalman               & Wavelets  & Outlier Detection \\                         
  \hline
   Oakland      & 72.21         & 75.92          & 63.44          &  22.92 \\ 
   San Diego    & 72.45        & 74.73           & 64.72         &   46.86\\ 
   Los Angeles  & 69.60        & 71.43          &   59.91       &   42.01\\ 
   Sacramento   & 68.42        & 70.85           & 59.30         &   43.79 \\ 
   Fresno       & 63.60        & 67.38           & 57.07         &  30.08 \\ 
   San Francisco & 66.14       & 67.14           & 55.85         &  38.01 \\ 
   Stockton     &  52.70        &  54.51        & 45.43         &  26.20\\ 
   Richmond     & 46.06       & 48.59          & 42.94          &  7.66\\ 
   Berkeley     & 38.25       & 40.07          & 35.48          &  13.13 \\ 
   \hline
\end{tabular}
\caption{Average sensitivity of spike identification methods for spikes of magnitudes ranging from 10-50\% increase over series mean}
\label{table:sensitivity}

\end{table}

\begin{table}[ht]
\centering
\begin{tabular}{p{2.5cm}p{2cm}p{2cm}p{2cm}p{2cm}}
  \hline 
  \textbf{City} &           & \textbf{Specificity}  &           &  \\ 
                &  ARIMA    & Kalman    & Wavelets  & Outlier Detection \\                         
  \hline
   Los Angeles  & 98.77        & 98.97       & 97.41         &  99.84 \\ 
   Oakland      & 98.68        & 98.84       & 97.03         &   99.54 \\
   San Diego    & 98.59        & 98.76       & 97.20         &  99.60 \\ 
   Fresno       & 98.28        & 98.73       & 97.55        &  99.65\\ 
   San Francisco &  98.56       & 98.68      & 97.23     &  99.76 \\ 
   Sacramento   & 98.37        & 98.56       & 97.22        & 99.51  \\ 
   Stockton     & 98.12        & 98.31       & 97.10        & 99.69 \\ 
   Richmond     & 97.89        & 98.14       & 96.85         &  99.79\\ 
   Berkeley     & 97.26        & 97.38       & 96.33        &  99.23\\ 

   \hline
\end{tabular}
\caption{Average specificity of spike identification methods for spikes of magnitudes ranging from 10-50\% increase over series mean}
\label{table:specificity}

\end{table}

\begin{table}[ht]
\centering
\begin{tabular}{p{2.5cm}p{2cm}p{2cm}p{2cm}p{2cm}}
  \hline 
  \textbf{City} &           & \textbf{Sensitivity}  &           &  \\ 
                &  ARIMA    & Kalman               & Wavelets  & Outlier Detection \\                         
  \hline
     San Diego    &   95.26        &  96.08           & 89.50          &    74.53\\ 
   Oakland      &   94.10        &  95.95           & 89.18          &  42.76 \\ 
   Los Angeles  &   94.16        &  95.01           & 87.20          &   70.90\\ 
   Sacramento   &   93.75        &  94.82           & 86.66          &   71.42\\ 
   San Francisco &  93.01        &  93.97           & 85.20          &  68.68 \\ 
   Fresno       &   89.67        &  91.57           & 83.77          &   50.63\\ 
   Stockton     &   85.07        &  86.18           & 76.12          &  52.19\\ 
   Richmond     &   78.49        &  80.10           & 74.78          &  17.00\\ 
   Berkeley     &   75.03        &  76.13           & 68.05          &  28.72\\ 
   \hline
\end{tabular}
\caption{Average sensitivity of spike identification methods for spikes with magnitude 50\% of series mean}
\label{table:sensitivity50}

\end{table}

\begin{table}[ht]
\centering
\begin{tabular}{p{2.5cm}p{2cm}p{2cm}p{2cm}p{2cm}}
  \hline 
  \textbf{City} &           & \textbf{Specificity}  &           &  \\ 
                &  ARIMA    & Kalman    & Wavelets  & Outlier Detection \\                         
  \hline
   Los Angeles  &  99.72         &  99.76           & 98.27          &  99.95 \\ 
   San Diego    &  99.64         &  99.69           & 98.18          &   99.86\\ 
   San Francisco & 99.60         &  99.63           & 98.12          &  99.92\\ 
   Oakland      &  99.53         &  99.61           & 98.20          &   99.64 \\ 
   Fresno       &  99.43         &  99.60           & 98.40          &  99.86 \\ 
   Sacramento   &  99.44         &  99.52           & 98.21          &   99.83\\ 
   Stockton     &  99.24         &  99.32           & 97.98          &  99.89\\ 
   Richmond     &  98.99         &  99.15           & 97.84          &  99.85\\ 
   Berkeley     &  98.26         &  98.37           & 97.30          &  99.59\\ 
   \hline
\end{tabular}
\caption{Average specificity of spike identification methods for spikes with magnitude 50\% of series mean}
\label{table:specificity50}

\end{table}

\begin{table}[ht]
  \centering
  \begin{tabular}{lc}
  \hline 
 \textbf{City}        & \textbf{Months with Spikes Detected}\\
 \hline
  Los Angeles         & 2005-07, 2006-07 \\
  Oakland             & 2007-03, 2009-09 \\
  San Diego           & 2005-07, 2007-05, 2007-10 \\ 
  Berkeley            & 2006-07, 2008-05, 2009-05, 2012-08 \\
  Fresno              & 2008-08, 2012-05 \\
  Sacramento          & 2009-08, 2010-08 \\
  San Francisco       & 2005-09, 2006-05, 2006-07, 2011-08, 2012-10 \\
  Stockton            & 2005-05, 2008-08 \\
  Richmond            & 2006-07, 2007-07, 2010-03, 2012-07 \\
  \hline
\end{tabular}
  \caption{Cities and months with violence spikes detected by the Kalman filter and smoother detection method}
  \label{table:spikes}
\end{table}

\newpage

\end{document}